\documentclass[journal]{IEEEtran}
\usepackage{cite}
\usepackage{amsmath}
\usepackage[table]{xcolor}

\ifCLASSINFOpdf
\else
   \usepackage[dvips]{graphicx}
\fi
\usepackage{url}

\usepackage{graphicx}

\begin{document}

\title{Speaker and Posture Classification using Instantaneous Intraspeech Breathing Features}

\author{At{\i}l \.Ilerialkan, Alptekin Temizel, \IEEEmembership{Member, IEEE}, H\"useyin Hac{\i}habibo\u{g}lu, \IEEEmembership{Senior Member, IEEE}
\thanks{Manuscript received May 25, 2020.}
\thanks{The authors are with the Graduate School of Informatics, Middle East Technical University (METU), Ankara, TR-06800, Turkey (e-mail: atil.ilerialkan@tai.com.tr, atemizel@metu.edu.tr, hhuseyin@metu.edu.tr).}
\thanks{© 2020 IEEE. Personal use of this material is permitted. Permission from IEEE must be obtained for all other uses, in any current or future media, including reprinting/republishing this material for advertising or promotional purposes, creating new collective works, for resale or redistribution to servers or lists, or reuse of any copyrighted component of this work in other works.}
}

\markboth{SUBMITTED FOR PUBLICATION TO IEEE SIGNAL PROCESSING LETTERS, VOL. tbd, NO. tbd, MAY 2020}
{Paper \MakeLowercase{\textit{et al.}}: for IEEE Journals}
\maketitle

\begin{abstract}
Acoustic features extracted from speech are widely used in problems such as biometric speaker identification and first-person activity detection. However, the use of speech for such purposes raises privacy issues as the  content is accessible to the processing party. In this work, we propose a method for speaker and posture classification using intraspeech breathing sounds. Instantaneous magnitude features are extracted using the Hilbert-Huang transform (HHT) and fed into a CNN-GRU network for classification of recordings from the open intraspeech breathing sound dataset, BreathBase, that we collected for this study. Using intraspeech breathing sounds, 87\% speaker classification and 98\% posture classification accuracy were obtained.
\end{abstract}


\IEEEpeerreviewmaketitle

\section{Introduction}
\IEEEPARstart{B}{reathing} interleaves speech. However, it is often considered to be a nuisance in speech processing since it does not contain semantic information. Breathing sounds depend on a variety of factors that depend on the speaker’s physiology such as an individual's lung capacity and the anatomy of their airway~\cite{gavriely2019breath}. They also vary depending on age, mass, current state and history of pathologic and physiological state~\cite{sovijarvi2000characteristics}. Therefore, it varies from person to person and is inimitable. In addition, it is universal in the sense that the breath signature of a person is largely independent of the spoken content and the language used. These properties make breathing sound a prominent candidate for use in biometric verification systems~\cite{lu2017sense}. 

There is a general lack of studies in the literature using intraspeech breathing in speaker identification tasks. An earlier pre-print~\cite{zhao2017speaker} aimed to demonstrate that breathing sounds can be used to identify speakers. However, it does not provide sufficient details of the algorithm and the employed dataset was not recorded in controlled environments. While a similar problem is tackled in this letter, we consider our work as a sequel to that study rather than a comparable one.

In this letter, we present a speaker and posture classification framework based on convolutional neural networks (CNN) and  gated  recurrent  units  (GRU) via the application of feature embedding using Hilbert-Huang transform (HHT). The underlying premise is the assumption that breathing signals are not stationary, meaning that commonly employed Fourier-based methods would not optimally represent breathing sounds. For that reason, we propose using instantaneous spectral features extracted from breathing sounds using HHT. We use time-varying instantaneous magnitudes as features to feed into the classifier. Results of an evaluation reveal that using the features extracted from signals captured by multiple microphones provide the best classification performance.

This letter is organized as follows. Background information about earlier research into the processing of intraspeech breathing sounds is given in Sec.~\ref{sect:bg}.  Sec.~\ref{sect:hht} presents the Hilbert-Huang transform. Sec.~\ref{sect:data} introduces the BreathBase dataset. Sec.~\ref{sect:clsf} introduces the features used in the study and presents the proposed classification approach. Sec.~\ref{sect:evalandconc} presents the evaluation results. Finally, Sec.~\ref{sect:concl} discusses the results and concludes the letter.

\section{Background} \label{sect:bg}
The detection and exact demarcation of human breath and its phases is a well-studied topic in speech signal processing. Among the earlier studies, mel-frequency \cite{igras2013wavelet, nakano2008analysis, ruinskiy2006algorithm, ruinskiy2007effective} and i-vector \cite{dumpala2017algorithm} based approaches are commonly used for feature extraction, while some other studies use raw breath sound signals~\cite{zhao2017speaker, sa2002automated, rosenwein2014detection}. A number of studies focus on particular problems and breathing types: classification of normal and abnormal breathing sounds is studied in \cite{cohen1984analysis}, breathing gestures (sniff, normal, and deep breathing) in \cite{chauhan2017breathprint} and phases (silence, breathing and snore) in \cite{karunajeewa2008silence}. As for speaker recognition, some studies use breathing sounds alone, while others include all non-speech sounds together with breath signals \cite{dumpala2017algorithm, zhao2017speaker, igras2013different, janicki2012impact, kinnunen2010overview, igras2016structure, lu2017sense}. Some of these earlier studies have focused on the detection and removal of breath signals and other non-speech sounds in order to improve the performance of automatic speech recognition (ASR)~\cite{rapcan2009automatic, zelasko2014hmm}. 

Datasets employed in these studies include recordings of ordinary participants as well as theatre and vocal artists. However, the employed data are rather inhomogenous in terms of the medium and recording conditions, including among others, television and telephone recordings \cite{nakano2008analysis, ruinskiy2006algorithm, ruinskiy2007effective, fukuda2011breath}. In order to collect these recordings, many different types of transducers were used, including different near or far-field microphones, imaging devices and respirators~\cite{gavriely2019breath}. 

Speech signals are considered to be stationary in short time intervals~\cite{deller1993discrete}.
However, non-stationary parts of the speech such as accents, emphases and different inflections could invalidate this assumption~\cite{tuske2012non}. Besides, methods relying on the assumption of stationarity, such as the widely used mel-frequency cepstral coefficients (MFCC), cannot accurately detect the events in intraspeech breathings that are sharply localized in time~\cite{abdalla2013dwt}.

\section{Hilbert-Huang Transform}
\label{sect:hht}
The Hilbert-Huang transform (HHT) is useful for performing time-frequency analysis on signals from non-stationary processes. The method is based on decomposing a signal with a finite time support into its intrinsic mode functions (IMF) that are recursively obtained from the signal itself, in contrast with transforms that use a fixed basis. HHT comprises two stages: empirical mode decomposition and Hilbert transform~\cite{huang1998empirical}.

Empirical mode decomposition (EMD) allows expressing a non-stationary signal as a linear combination of its intrinsic mode functions (IMF) from which instantaneous frequencies can be calculated. Using EMD, a non-stationary signal, $x_p(n)$ with a finite time support in $0\leq{}n<N_p$ can be expressed as a linear combination of its $K_p$ IMFs, $c_{p,k}(n)$, and a residual signal, $r_{p}(n)$, that typically has very low energy. All IMFs as well as the residual have the same time support as the original signal, such that:
\begin{equation}
    x_p(n)=\sum_{k=1}^{K_p}c_{p,k}(n) + r_{p}(n).
\end{equation}
The IMFs should satisfy the following properties in order to be useful in the calculation of instantaneous spectrum: i) the number of extrema and the number of zero crossings should differ at most by one and ii) the mean of the envelopes defined by the maxima and minima, respectively, should be zero~\cite{huang1998empirical}. EMD extracts these IMFs by alternating between identification and inversion of the signal envelope and recursive extraction of IMFs from $x_p(n)$ until either the residual energy is below a predefined threshold or a predefined number of iterations is completed. IMFs with a lower index have higher frequency content.

Extraction of instantaneous spectrum requires the computation of the discrete Hilbert transform of each IMF to obtain their analytic counterparts given as:
\begin{equation}
    z_{p,k}(n)=c_{p,k}(n)+j\mathcal{H}\left\{c_{p,k}(n)\right\}=a_{p,k}(n)e^{j\Psi_{p,k}(n)}
\end{equation}
where the discrete Hilbert transform can be calculated using the DFT of the signal as:
\begin{equation}
    \mathcal{H}\left\{x(n)\right\}=\frac{1}{N}\sum_{k=0}^{N-1}\left[H(k)\sum_{n=0}^{N-1}x(n)e^{-j\frac{2\pi{}}{N}nk}\right]e^{j\frac{2\pi{}}{N}nk}.
\end{equation}
Here, $H(k)=2\left[u(k)-u(k-\frac{N}{2}-1)\right]-\delta{}(k)-\delta{}(k-\frac{N}{2})$ with $u(.)$ is a unit step sequence and $\delta(k)$ is a unit sample, selects the positive frequency components associated with the quadrature part of the signal. Note that an analytic signal is calculated for each IMF obtained by using EMD.

Depending on the number of IMFs obtained, HHT typically results in a sparse intantaneous spectrum that comprises instantaneous magnitude and instantaneous frequency calculated at each time instant. The feature embedding we employ in this letter depends on the instantaneous magnitude that can be calculated from the analytic counterpart of each IMF, defined as:
\begin{equation}
    \mathcal{M}_{p,k}(n)=[z_{p,k}(n)z_{p,k}^*(n)]^{1/2},
\end{equation}
where $(\cdot)^{*}$ represents complex conjugation.

\section{Data Collection} \label{sect:data}

An important shortcoming with the earlier studies on speaker classification using intraspeech breathing sounds is the lack of a dataset comprising samples recorded under strictly controlled conditions. To address this problem, we developed an open dataset of intraspeech breathing sounds called BreathBase~\cite{breathbase}.

\subsection{Participants and Recording Procedure}
A total of 20 participants, who are active licensed American Football players with good health, took part in the data collection. An equal number of male and participants who are 20 to 30 years old participated in data collection. 

Each participant was requested to read one and a half pages long texts from \textit{The Sorrows of Young Werther} of Johann Wolfgang von Goethe and \textit{The Metamorphosis} of Franz Kafka. The sequences of the original texts were randomized and auto-translated into Turkish using Google Translator to distort their meanings. In order to avoid the occurrence of pauses like silences, emphasis and filled pauses (like `umm' voices) while reading, the entire text was decapitalized and all punctuation marks were removed.

Recordings were collected at the METU Spatial Audio Research Group (SPARG) Lab which is a specially designed acoustic chamber with very low reverberation ($T_{60}=80$ ms). The chamber has no parallel walls, including its floor and ceiling, in order to eliminate standing waves. The peak background noise level was measured to be 40 dB SPL. 

\subsection{Recording Setup}
The recording setup comprised four microphones (three R\o{}de\textsuperscript{TM} M5 and one DPA\textsuperscript{TM} 4060). DPA\textsuperscript{TM} 4060, which is an omnidirectional subminiature microphone, was placed next to the speaker's mouth. R\o{}de\textsuperscript{TM} M5 microphones, which are cardioid microphones matched for their frequency responses, were positioned in such a way that for each recorded posture, there was one microphone that was at a 1.5 m line-of-sight distance from the speaker's mouth. The audio interface used in the recordings was a MOTU 896mk3 Hybrid multichannel sound card with low-noise, linear-phase microphone preamplifiers.

\subsection{Dataset}
BreathBase provides tagging for a total of 5070 intraspeech breath signals for 5 different postures (\textit{high sitting}, \textit{low sitting}, \textit{standing}, \textit{standing with hands behind head} and \textit{lying}) using 4 microphones. These postures were selected not only for their practical relevance (e.g.~in assisted living technologies) but also based on the presupposition that these postures would result in distinctly different conditions in terms of speech generation, and particularly for the generation of intraspeech breath sounds. For example, the low sitting condition squeezes the diaphragm and decreases the lung volume, changing not only the features of speech but also of intraspeech breath sounds.

The dataset contains a minimum of 89, a maximum of 710 and an average of 254 intraspeech breath instances per participant. Intraspeech breath signals were extracted by frame-based level thresholding. Low-frequency noise observed at the cardioid microphones due to the proximity effect~\cite{kleiner2013electroacoustics} was reduced by using a 4097-tap high-pass, linear-phase FIR filter with a cutoff frequency of 70 Hz. The same filter was also applied to the omnidirectional subminiature microphone to mitigate the risk of classifiers fitting to the microphone properties. Each breath signal was time aligned via cross-correlation and normalized for its energy to prevent classifiers from learning the time delay of arrival and the level differences between the microphones, respectively. 

In order to test our assumption about the non-stationarity of breath signals, we used the augmented Dickey-Fuller (ADF) test~\cite{greene2003econometric} on breath samples. The null hypothesis for the ADF test is that the time series comes from a non-stationary process. The ADF test, as applied to all 5070 samples in our dataset revealed that the null hypothesis could be rejected at $p=0.01$ level only for $783$ breath samples. While this result by itself does not guarantee that the other samples in the dataset are non-stationary, it is a strong indication in that direction. Therefore, feature embedding using HHT instead of using Fourier-based methods is practically justified.

\section{Features and Classifier Architectures} \label{sect:clsf}
\subsection{Features}
The proposed method uses instantaneous magnitudes extracted using the Hilbert-Huang transform (HHT) which are fed into the classifier architectures in various configurations. In all cases, a fixed number of $K_p=K=9$ IMFs are extracted from intraspeech breath recordings for $Q=4$ channels. Let us represent the instantaneous magnitude vector comprising instantaneous magnitudes obtained using HHT from channel $c$, for breath instance $p$ as: 
\begin{equation}
\mathbf{m}_p
^{(c)}(n)=[\mathcal{M}_{p,1}
^{(c)}(n)\ \mathcal{M}_{p,2}^{(c)}(n)\ \cdots{}\mathcal{M}_{p,K}^{(c)}(n)]
^T.
\end{equation}
Using this $K\times{}1$ vector as a unit, several multivariate time series were constructed for use in classification of postures and speakers. Time index is omitted henceforth for notational brevity and the length of each time series will implicitly be equal to the time support, $N_p$, of the individual breath instance, $p$. The configurations employed in this letter are as follows:
\begin{enumerate}
    \item \textbf{Channel-0:} This configuration uses data only from the subminiature microphone positioned close to subjects' mouths and mainly serves as a baseline for the other tested configurations. Each sample of the multivariate time series used for representing a single breath instance is given as a $K\times{}1$ vector such that:
    \begin{equation}
        \mathbf{M}_p^{(0)}=\mathbf{m}_p^{(0)}.
    \end{equation}
    \item \textbf{Split channel:} This configuration uses feature vectors obtained from all microphones, i.e.~$\{\mathbf{M}_p^{(m)}=\mathbf{m}_p^{(m)}\}_{m=1\cdots{}4}$ resulting in four times more data to train the classifier networks. 
     \item \textbf{All channels (Ordered):} This configuration extends the feature vectors vertically, resulting in a $4K\times{}1$ vector at each time instant such that:
    \begin{equation}
        \mathbf{M}_{p, \textit{ordered}}^T=[\mathbf{m}_p^{(0),T}\ \mathbf{m}_p^{(1),T}\ \mathbf{m}_p^{(2), T}\ \mathbf{m}_p^{(3), T}]
    \end{equation}
    This way, the ordering of different microphones is preserved in each feature vector input to the networks.
    \item \textbf{All channels (Shuffled)}: This configuration randomizes the channel ordering for each sample of the previous configuration such that:
    \begin{equation}
        \mathbf{M}_{p, \textit{shuffled}}=\mathbf{B}\mathbf{M}_{p, \textit{ordered}}
    \end{equation}
    where $\mathbf{B}=\mathbf{P}\otimes\mathbf{I}_{K\times{}K}$ is a block permutation matrix formed by the Kronecker product of a $4\times{}4$ random permutation matrix $\mathbf{P}$ with a $K\times{}K$ identity matrix, $\mathbf{I}_{K\times{}K}$. Notice that, in order to prevent adaptation to a single permutation, $\mathbf{P}$ is randomly updated for each sample. 
\end{enumerate}

\subsection{Classifier Architectures}
Since breathing sound is a continuous, time-series data, we anticipate that the feature vectors of neighboring time steps convey strong relations in-between. Hence a deep convolutional network has been used to exploit these correlations in a hierarchical manner. Gated Recurrent Units (GRU) \cite{cho2014learning} were used to exploit correlations in time, based on already dense high-level features extracted by the convolution layers, to support sequence classification. GRU, while being a simpler architecture using fewer gates, provides similar performance to long short-term memory (LSTM) networks.

\begin{figure}
\centering
\includegraphics[keepaspectratio, width=.4\textwidth]{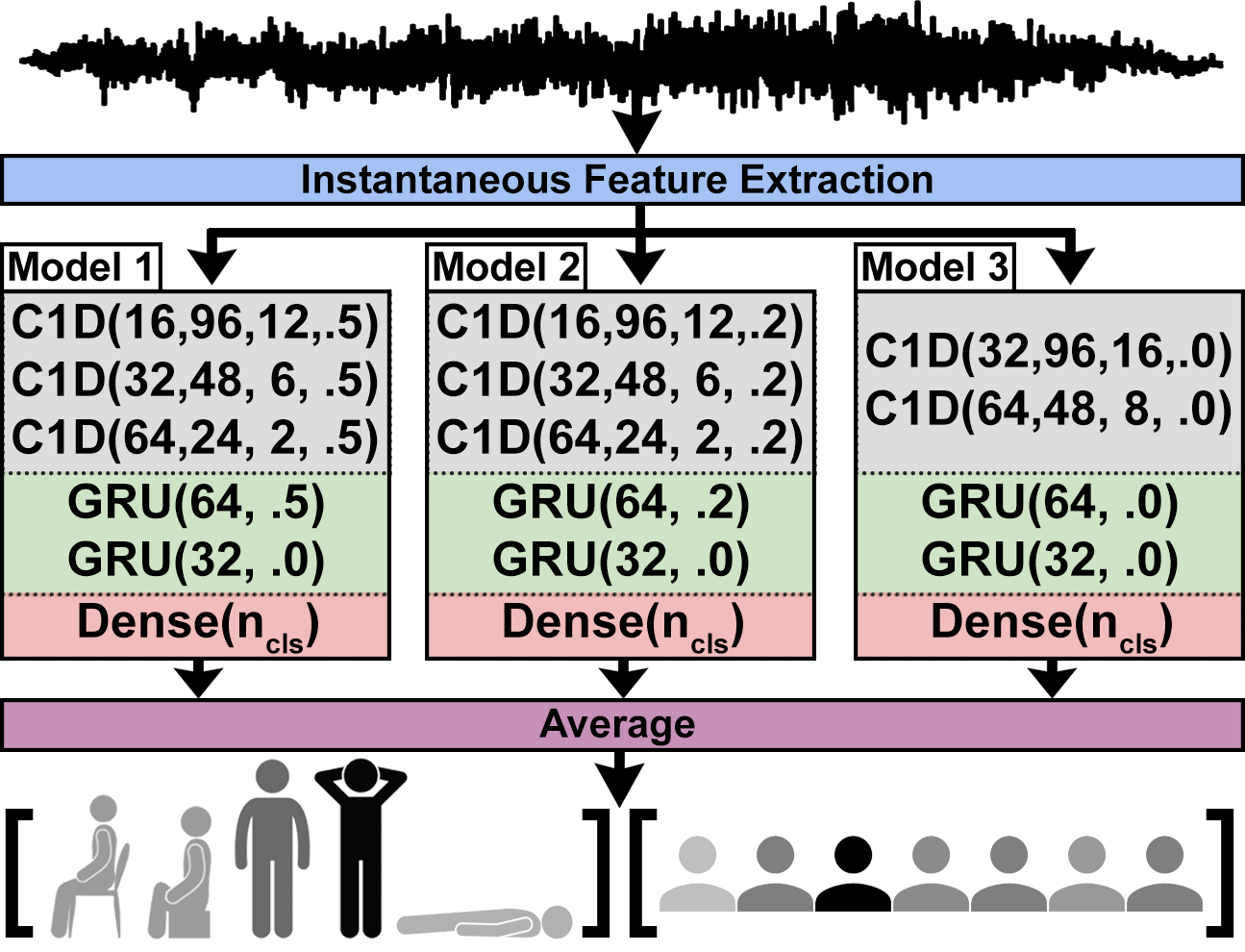}
\caption{Ensembled classifier network architecture. 1-dimensional convolution layer with rectified linear unit (ReLU) activation is formulated as: \textit{C1D(number of filters, kernel size, stride, drop-out rate)}. Gated Recurrent Unit layer with hyperbolic tangent (tanh) activation is formulated as: \textit{GRU(number of units, drop-out rate)}. Dense layer with softmax activation is formulated as:  \textit{Dense(number of nodes)}.}
\label{fig:architectures}
\end{figure}

Three different models have been developed for classification. The models use stride values smaller than the kernel lengths, providing overlapping on data and preventing sharp filter boundaries. The architectures use varying amounts of dropout rates to prevent overfitting. Additionally, an ensemble network has been created using these three classifiers. Details of the employed network layers are shown in Fig.~\ref{fig:architectures}.

\begin{table*}[!t]
\centering
\caption{Results for 3-posture and 5-posture and speaker classification using instantaneous magnitude features with different network architectures and different channel modes.}
\label{table:Results}
\setlength\tabcolsep{4pt} 
\begin{tabular}{|l|c|c|c|c|c|c|c|c|c|c|c|c|}
\hline
                              & \multicolumn{4}{c|}{\textbf{\textit{3-posture}}}                  & \multicolumn{4}{c|}{\textbf{\textit{5-posture}}}                  & \multicolumn{4}{c|}{\textbf{\textit{Speaker}}}                  \\ \hline
\textbf{\textit{Channel Mode}}         & \textbf{Model 1}      & \textbf{Model 2} & \textbf{Model 3} & \textbf{Ens.} & \textbf{Model 1} & \textbf{Model 2} & \textbf{Model 3}      & \textbf{Ens.} & \textbf{Model 1} & \textbf{Model 2} & \textbf{Model 3} & \textbf{Ens.}    \\ \hline
\textbf{Channel-0}                    & 0.51          & 0.47     & 0.47     & 0.49       & 0.36     & 0.31     & 0.33          & 0.36       & 0.61     & 0.56     & 0.60     & 0.63          \\ \hline
\textbf{Split channel}                & 0.62          & 0.60     & 0.57     & 0.63       & 0.44     & 0.43     & 0.41          & 0.45       & 0.68     & 0.68     & 0.65     & 0.72          \\ \hline
\textbf{All-channels (Ordered)}        & \textbf{0.98} & \textbf{0.98}     & \textbf{0.98}     & \textbf{0.98}       & 0.82     & 0.81     & \textbf{0.87} & 0.85       & 0.83     & 0.78     & 0.83     & \textbf{0.87} \\ \hline
\textbf{All-channels (Shuffled)} & 0.85          & 0.65     & 0.74     & 0.86       & 0.48     & 0.52     & 0.54          & 0.55       & 0.68     & 0.65     & 0.69     & 0.74          \\ \hline

\end{tabular}
\end{table*}

\section{Experimental Evaluation} \label{sect:evalandconc}
 We classified the recordings both by speaker identity and by recording posture. Posture classification is done in two different settings: i) considering each of the 5 original recording postures a separate class,  ii) grouping them into 3 classes which are more general (\textit{sitting}, \textit{standing} and \textit{lying}) by merging \textit{high sitting} and \textit{low sitting} postures under \textit{sitting} and \textit{standing} and \textit{standing with hands behind head} postures under \textit{standing}. Train/test split is 80/20 and done in a stratified manner.  

The results are summarized in Table \ref{table:Results} for different channel modes and using different network architectures for 3-posture, 5-posture and speaker classification problems. Channel-0 mode was used as a benchmark to demonstrate the effect of using multiple channels and provided accuracies of 51\%, 36\% and 63\% for 3-posture\ 5-posture and speaker classification, respectively using the ensembled network. Better accuracy levels were achieved when all channels are used to train and test the classifiers either sequentially (i.e.
~split channel mode) or in parallel (i.e.~ two all-channels modes). Among the modes that all channels are used, \textit{All Channels (Ordered)} was the best performing configuration for all cases. For 3-posture classification, all three models and the ensembled network gives similar results with a 98\% accuracy. For 5-posture classification, Model-3 performs the best with 87\% accuracy. For speaker classification, the ensembled network has the best performance with 87\% accuracy.

\begin{figure}[!t]
\centering
\includegraphics[keepaspectratio,width=.46\textwidth]{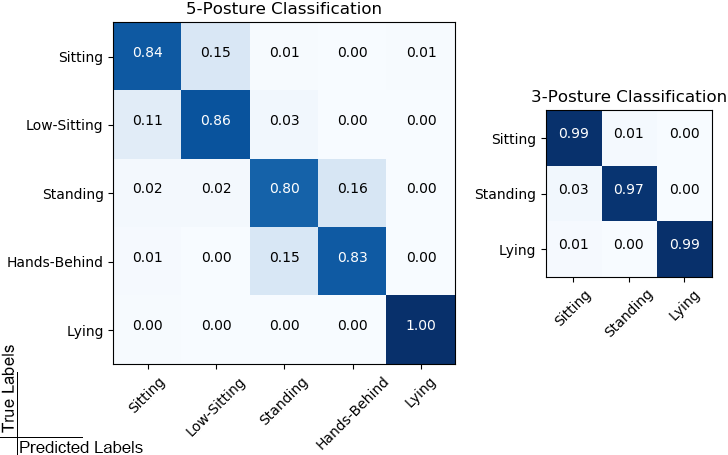}
\caption{Confusion matrices for 3 and 5-posture class recognition for the best performing models.}
\label{fig:confmat}
\end{figure}

The confusion matrix for the 5-posture task in Fig.~\ref{fig:confmat} shows that the two standing and sitting positions were confused within themselves in this setting. The confusion matrix for the 3-posture task shows that this confusion was resolved when these positions were merged into the more general sitting and standing classes. The lying position, on the other hand, was classified distinctly apart from other positions.

\section{Conclusions} \label{sect:concl}
We proposed a method for speaker and posture classification using only intraspeech breathing sound information in this letter. We also developed the BreathBase dataset which includes 5070 intraspeech breathing signals for this purpose. To the best of our knowledge, BreathBase is currently the only intraspeech breath sound dataset.

Since the analysis of the breath instances revealed strong indications that intraspeech breath signals are non-stationary, instantaneous magnitude features extracted from the Hilbert-Huang transform are used to classify them. The approach we propose, takes into account the temporal structure of instantaneous magnitudes extracted using HHT from four microphones. We implemented multiple variations of CNN-GRU networks to investigate the potential of the proposed feature embedding. Our experiments show the potential of the proposed method which achieves 87\% speaker recognition accuracy in 20 speaker classes and with 98\% and 87\% posture recognition accuracy in 3-posture and 5-posture classes, respectively. An important finding is that using multiple simultaneous recordings improved the accuracy, especially for posture classification.

We aimed to make sure that the networks we employed learned postures and speakers, and not, for example, the microphone characteristics. The breath instances were normalized and time-aligned for that purpose and three of the microphones had matched frequency responses. However, it is possible that such adaptation might have occured for the \textit{All Channels (Ordered)} configuration in posture classification since each posture corresponds also to a certain spatial configuration of microphones. Still, since the \textit{All Channels (Shuffled)} configuration also provides good accuracy, it is likely that the classfiers did not learn microphone characteristics, or even if they did, this effect was not substantial. Similarly, \textit{All Channels (Ordered)} configuration provided high accuracy also for the speaker identification task where there is no risk of learning the microphone characteristics since recordings for each speaker were made using at all the different microphone configurations.

\bibliographystyle{IEEEtran}
\bibliography{IEEEabrv, paper}

\begin{thebibliography}{10}
\providecommand{\url}[1]{#1}
\csname url@samestyle\endcsname
\providecommand{\newblock}{\relax}
\providecommand{\bibinfo}[2]{#2}
\providecommand{\BIBentrySTDinterwordspacing}{\spaceskip=0pt\relax}
\providecommand{\BIBentryALTinterwordstretchfactor}{4}
\providecommand{\BIBentryALTinterwordspacing}{\spaceskip=\fontdimen2\font plus
\BIBentryALTinterwordstretchfactor\fontdimen3\font minus
  \fontdimen4\font\relax}
\providecommand{\BIBforeignlanguage}[2]{{%
\expandafter\ifx\csname l@#1\endcsname\relax
\typeout{** WARNING: IEEEtran.bst: No hyphenation pattern has been}%
\typeout{** loaded for the language `#1'. Using the pattern for}%
\typeout{** the default language instead.}%
\else
\language=\csname l@#1\endcsname
\fi
#2}}
\providecommand{\BIBdecl}{\relax}
\BIBdecl

\bibitem{gavriely2019breath}
N.~Gavriely and D.~W. Cugall, \emph{Breath sounds methodology}.\hskip 1em plus
  0.5em minus 0.4em\relax CRC Press, 2019.

\bibitem{sovijarvi2000characteristics}
A.~Sovijärvi, L.~Malmberg, G.~Charbonneau, J.~Vanderschoot, F.~Dalmasso,
  C.~Sacco, M.~Rossi, and J.~Earis, ``Characteristic of breath sounds and
  adventitious respiratory sounds,'' \emph{Eur Respir Rev}, vol.~10, pp.
  591--596, 01 2000.

\bibitem{lu2017sense}
L.~Lu, L.~Liu, M.~J. Hussain, and Y.~Liu, ``I sense you by breath: Speaker
  recognition via breath biometrics,'' \emph{{IEEE} Trans. Depend. Sec.
  Comput.}, vol.~17, no.~2, pp. 306 -- 319, October 2017.

\bibitem{zhao2017speaker}
W.~{Zhao}, Y.~{Gao}, and R.~{Singh}, ``{Speaker identification from the sound
  of the human breath},'' \emph{arXiv e-prints}, p. arXiv:1712.00171, Nov.
  2017.

\bibitem{igras2013wavelet}
M.~Igras and B.~Zi{\'o}lko, ``Wavelet method for breath detection in audio
  signals,'' in \emph{Proc. 2013 IEEE Int. Conf. Multimedia and Expo
  (ICME)}.\hskip 1em plus 0.5em minus 0.4em\relax IEEE, July 2013, pp. 1--6.

\bibitem{nakano2008analysis}
T.~Nakano, ``Analysis and automatic detection of breath sounds in unaccompanied
  singing voice,'' \emph{Proc. 10th Int. Conf. Music Perception Cognition
  (ICMPC10)}, pp. 387--390, 2008.

\bibitem{ruinskiy2006algorithm}
D.~Ruinskiy and Y.~Lavner, ``An algorithm for accurate breath detection in
  speech and song signals,'' in \emph{Proc. 2006 IEEE 24th Conv. Electrical
  Electronics Engineers Israel}.\hskip 1em plus 0.5em minus 0.4em\relax IEEE,
  Nov. 2006, pp. 315--319.

\bibitem{ruinskiy2007effective}
------, ``An effective algorithm for automatic detection and exact demarcation
  of breath sounds in speech and song signals,'' \emph{{IEEE} Trans. Audio,
  Speech, Language Process.}, vol.~15, no.~3, pp. 838--850, 2007.

\bibitem{dumpala2017algorithm}
S.~H. Dumpala and K.~R. Alluri, ``An algorithm for detection of breath sounds
  in spontaneous speech with application to speaker recognition,'' in
  \emph{Proc. 2017 Int. Conf. on Speech and Computer (SPECOM 2017)}.\hskip 1em
  plus 0.5em minus 0.4em\relax Springer, 2017, pp. 98--108.

\bibitem{sa2002automated}
R.~C. S{\'a} and Y.~Verbandt, ``Automated breath detection on long-duration
  signals using feedforward backpropagation artificial neural networks,''
  \emph{{IEEE} Trans. Biomed. Eng.}, vol.~49, no.~10, pp. 1130--1141, 2002.

\bibitem{rosenwein2014detection}
T.~Rosenwein, E.~Dafna, A.~Tarasiuk, and Y.~Zigel, ``Detection of breathing
  sounds during sleep using non-contact audio recordings,'' in \emph{2014 36th
  Annu. Int. Conf. IEEE Engineering Medicine and Biology Society}.\hskip 1em
  plus 0.5em minus 0.4em\relax IEEE, Nov. 2014, pp. 1489--1492.

\bibitem{cohen1984analysis}
A.~Cohen and D.~Landsberg, ``Analysis and automatic classification of breath
  sounds,'' \emph{{IEEE} Trans. Biomed. Eng.}, vol.~31, no.~9, pp. 585--590,
  Sep. 1984.

\bibitem{chauhan2017breathprint}
J.~Chauhan, Y.~Hu, S.~Seneviratne, A.~Misra, A.~Seneviratne, and Y.~Lee,
  ``{BreathPrint}: Breathing acoustics-based user authentication,'' in
  \emph{Proc. 15th Ann. Int. Conf. Mobile Syst., Appl. Services (MobiSys
  '17)}.\hskip 1em plus 0.5em minus 0.4em\relax ACM, June 2017, pp. 278--291.

\bibitem{karunajeewa2008silence}
A.~S. Karunajeewa, U.~R. Abeyratne, and C.~Hukins, ``Silence--breathing--snore
  classification from snore-related sounds,'' \emph{Physiol. Meas.}, vol.~29,
  no.~2, p. 227, Feb. 2008.

\bibitem{igras2013different}
M.~Igras and B.~Zi{\'o}{\l}ko, ``Different types of pauses as a source of
  information for biometry,'' in \emph{Proc. 8th Int. Work. Models Anal. Vocal
  Emissions for Biomedical Appl.}, Dec. 2013, pp. 197--200.

\bibitem{janicki2012impact}
A.~Janicki, ``On the impact of non-speech sounds on speaker recognition,'' in
  \emph{Proc. 15th Int. Conf. Text, Speech and Dialogue}.\hskip 1em plus 0.5em
  minus 0.4em\relax Springer, Sep. 2012, pp. 566--572.

\bibitem{kinnunen2010overview}
T.~Kinnunen and H.~Li, ``An overview of text-independent speaker recognition:
  From features to supervectors,'' \emph{Speech Comm.}, vol.~52, no.~1, pp.
  12--40, Jan. 2010.

\bibitem{igras2016structure}
M.~Igras-Cybulska, B.~Zi{\'o}{\l}ko, P.~{\.Z}elasko, and M.~Witkowski,
  ``Structure of pauses in speech in the context of speaker verification and
  classification of speech type,'' \emph{EURASIP J. Audio, Speech, and Music
  Process.}, vol. 2016, no.~1, p.~18, Jan. 2016.

\bibitem{rapcan2009automatic}
V.~Rapcan, S.~D'Arcy, and R.~B. Reilly, ``Automatic breath sound detection and
  removal for cognitive studies of speech and language,'' in \emph{Proc. 2009
  IET Irish Signals and Syst. Conf. (ISSC-2009)}.\hskip 1em plus 0.5em minus
  0.4em\relax IET, June 2009.

\bibitem{zelasko2014hmm}
P.~{\.Z}elasko, T.~Jadczyk, and B.~Zi{\'o}{\l}ko, ``{HMM}-based breath and
  filled pauses elimination in {ASR},'' in \emph{Proc. 11th Int. Conf Signal
  Process. and Multimedia Appl. (SIGMAP-2014)}, Aug. 2014, pp. 255--260.

\bibitem{fukuda2011breath}
T.~Fukuda, O.~Ichikawa, and M.~Nishimura, ``Breath-detection-based telephony
  speech phrasing,'' in \emph{Proc. 12th Ann. Conf. Int. Speech Comm. Assoc.
  (INTERSPEECH 2011)}, Aug. 2011, pp. 2625--2628.

\bibitem{deller1993discrete}
J.~R. Deller~Jr, J.~G. Proakis, and J.~H. Hansen, \emph{Discrete time
  processing of speech signals}.\hskip 1em plus 0.5em minus 0.4em\relax
  Prentice Hall PTR, 1993.

\bibitem{tuske2012non}
Z.~T{\"u}ske, F.~R. Drepper, and R.~Schl{\"u}ter, ``Non-stationary signal
  processing and its application in speech recognition,'' in \emph{Proc.
  SAPA-SCALE Conference}, Sep. 2012, pp. 34--39.

\bibitem{abdalla2013dwt}
M.~I. Abdalla, H.~M. Abobakr, and T.~S. Gaafar, ``{DWT} and {MFCCs} based
  feature extraction methods for isolated word recognition,'' \emph{Int. J.
  Comp. Appl.}, vol.~69, no.~20, May 2013.

\bibitem{huang1998empirical}
N.~E. Huang, Z.~Shen, S.~R. Long, M.~C. Wu, H.~H. Shih, Q.~Zheng, N.-C. Yen,
  C.~C. Tung, and H.~H. Liu, ``The empirical mode decomposition and the
  {Hilbert} spectrum for nonlinear and non-stationary time series analysis,''
  \emph{Proc. Royal Soc. London Series A: Math. Phys. Eng. Sci.}, vol. 454, no.
  1971, pp. 903--995, Mar. 1998.

\bibitem{breathbase}
\BIBentryALTinterwordspacing
A.~İlerialkan, ``{BreathBase: Intra-Speech Breathing Dataset},'' May 2020.
  [Online]. Available: \url{https://doi.org/10.5281/zenodo.3841039}
\BIBentrySTDinterwordspacing

\bibitem{kleiner2013electroacoustics}
M.~Kleiner, \emph{Electroacoustics}.\hskip 1em plus 0.5em minus 0.4em\relax CRC
  Press, 2013.

\bibitem{greene2003econometric}
W.~H. Greene, \emph{Econometric Analysis}.\hskip 1em plus 0.5em minus
  0.4em\relax Prentice Hall, New Jersey, 2003.

\bibitem{cho2014learning}
K.~{Cho}, B.~{van Merrienboer}, C.~{Gulcehre}, D.~{Bahdanau}, F.~{Bougares},
  H.~{Schwenk}, and Y.~{Bengio}, ``{Learning Phrase Representations using RNN
  Encoder-Decoder for Statistical Machine Translation},'' \emph{arXiv
  e-prints}, p. arXiv:1406.1078, Jun. 2014.

\end{thebibliography}

\end{document}